\documentclass[prl,reprint,amsmath,amssymb,aps,preprintnumbers,longbibliography]{revtex4-1}

\usepackage[colorlinks=true,linkcolor=black, citecolor=black,
urlcolor=black]{hyperref}

\usepackage{multirow,graphics}
\usepackage{enumerate}
\usepackage{amstext}
\usepackage{amssymb}
\usepackage{amsmath}
\usepackage{graphicx}
\usepackage{color}
\usepackage{tikz}
\usepackage{psfrag}
\usepackage{soul}

\usepackage{thumbpdf}
\usepackage{esint,shuffle,psfrag}
 \usepackage[utf8]{inputenc}

\allowdisplaybreaks
\usepackage{varioref}
\usepackage{amsmath,amsfonts,amsthm,mathrsfs}
\usepackage{enumerate}
\usepackage{verbatim}
\usepackage{wrapfig}
\usepackage{appendix}
\usepackage{amstext}
\usepackage{amssymb}
\usepackage{graphicx}
\usepackage{color}
\usepackage{varioref}
\usepackage{multirow,graphics}
\usepackage{epstopdf}
\usepackage{bbm}
\usepackage{slashed}
\usepackage{psfrag}
\usepackage{relsize}
\usepackage{bbm}
\usepackage{mathrsfs}
\usepackage{epsfig}
\usepackage{upgreek}

\usepackage{mathtools}
       \usepackage{dsfont}

\setcitestyle{square,numbers}

\begin{document}

\newcommand{\IM}{{\rm Im}\,}
\newcommand{\card}{\#}
\newcommand{\la}[1]{\label{#1}}
\newcommand{\eq}[1]{(\ref{#1}),} 
\newcommand{\figref}[1]{Fig. \ref{#1}}
\newcommand{\abs}[1]{\left|#1\right|}
\newcommand{\comD}[1]{{\color{red}#1\color{black}}}
\newcommand{\p}{{\partial}}
\newcommand{\Tr}{{\text{Tr}}}
\newcommand{\tr}{{\text{tr}}}
\newcommand{\sym}{${\cal N}=4$ SYM becomes a non-unitary and non-supersymmetric CFT. }
\newcommand{\como}[1]{{\color[rgb]{0.0,0.1,0.9} {\bf \"O:} #1} }

\makeatletter
\newcommand{\subalign}[1]{%
  \vcenter{%
    \Let@ \restore@math@cr \default@tag
    \baselineskip\fontdimen10 \scriptfont\tw@
    \advance\baselineskip\fontdimen12 \scriptfont\tw@
    \lineskip\thr@@\fontdimen8 \scriptfont\thr@@
    \lineskiplimit\lineskip
    \ialign{\hfil$\m@th\scriptstyle##$&$\m@th\scriptstyle{}##$\crcr
      #1\crcr
    }%
  }
}
\makeatother

\newcommand{\mzvv}[2]{
  \zeta_{
    \subalign{
      &#1,\\
      &#2
    }
}
}

\newcommand{\mzvvv}[3]{
  \zeta_{
    \subalign{
      &#1,\\
      &#2,\\
      &#3
    }
}
  }

\makeatletter
     \@ifundefined{usebibtex}{\newcommand{\ifbibtexelse}[2]{#2}} {\newcommand{\ifbibtexelse}[2]{#1}}
\makeatother

\preprint{}


\usetikzlibrary{decorations.pathmorphing}
\usetikzlibrary{decorations.markings}
\usetikzlibrary{intersections}
\usetikzlibrary{calc}

\tikzset{
photon/.style={decorate, decoration={snake}},
particle/.style={postaction={decorate},
    decoration={markings,mark=at position .5 with {\arrow{>}}}},
antiparticle/.style={postaction={decorate},
    decoration={markings,mark=at position .5 with {\arrow{<}}}},
gluon/.style={decorate, decoration={coil,amplitude=2pt, segment length=4pt},color=purple},
wilson/.style={color=blue, thick},
scalarZ/.style={postaction={decorate},decoration={markings, mark=at position .5 with{\arrow[scale=1]{stealth}}}},
scalarX/.style={postaction={decorate}, dashed, dash pattern = on 4pt off 2pt, dash phase = 2pt, decoration={markings, mark=at position .53 with{\arrow[scale=1]{stealth}}}},
scalarZw/.style={postaction={decorate},decoration={markings, mark=at position .75 with{\arrow[scale=1]{stealth}}}},
scalarXw/.style={postaction={decorate}, dashed, dash pattern = on 4pt off 2pt, dash phase = 2pt, decoration={markings, mark=at position .60 with{\arrow[scale=1]{stealth}}}}
}

 \newcommand{\doublewheelsmall}{
   \begin{minipage}[c]{1cm}
     
     \begin{center}
       \begin{tikzpicture}[scale=0.3]
         \foreach \m in {1,2} {
           \draw (0.75*\m,0) arc[radius = 0.75*\m,start angle = 0, end angle = 300] ;
           \draw[black, densely dotted] (300:0.78*\m) arc[radius = 0.75*\m,start angle = -60, end angle = 0];
         }
         \foreach \t in {1,2,...,5} {
           \draw (0,0) -- (60*\t:1.5);
         }
         \draw[black,densely dotted] (0,0) -- (0:1.5);
       \end{tikzpicture}
     \end{center}
   \end{minipage}
 }


\newcommand{\footnoteab}[2]{\ifbibtexelse{%
\footnotetext{#1}%
\footnotetext{#2}%
\cite{Note1,Note2}%
}{%
\newcommand{\textfootnotea}{#1}%
\newcommand{\textfootnoteab}{#2}%
\cite{thefootnotea,thefootnoteab}}}

\definecolor{green(pigment)}{rgb}{0.0, 0.65, 0.31}
\def\e{\epsilon}
     \def\bT{{\bf T}}
    \def\bQ{{\bf Q}}
    \def\wT{{\mathbb{T}}}
    \def\wQ{{\mathbb{Q}}}
    \def\ttQ{{\bar Q}}
    \def\tQ{{\tilde \bP}}
        \def\bP{{\bf P}}
        \def\dq{{\dot q}}
    \def\CF{{\cal F}}
    \def\cC{\CF}
    
     \def\l{\lambda}
\def\hbZ{{\widehat{ Z}}}
\def\bZ{{\resizebox{0.28cm}{0.33cm}{$\hspace{0.03cm}\check {\hspace{-0.03cm}\resizebox{0.14cm}{0.18cm}{$Z$}}$}}}

\definecolor{forestgreen(traditional)}{rgb}{0.0, 0.27, 0.13}

\title{Stampedes II: Null Polygons in Conformal Gauge Theory}

\author{Enrico Olivucci$^{a}$ and\, Pedro Vieira$^{a,b}$}

\affiliation{%
\\ \\
\({}^{a}\) Perimeter Institute for Theoretical Physics, Waterloo, Ontario N2L 2Y5, Canada
 \\
\(^{b}\) 
ICTP South American Institute for Fundamental Research, IFT-UNESP, S\~ao Paulo, SP Brazil 01440-070 
\\
}

\begin{abstract}

We consider correlation functions of single trace operators approaching the cusps of null polygons in a double-scaling limit where 
so-called \emph{cusp times} $t_i^2 = g^2 \log x_{i-1,i}^2\log x_{i,i+1}^2$ are held fixed and the t'Hooft coupling is small. With the help of stampedes, symbols and educated guesses, we find that any such correlator can be uniquely fixed through a set of coupled lattice PDEs of Toda type with several intriguing novel features. These results hold for most conformal gauge theories with a large number of colours, including planar $\mathcal{N}=4$ SYM. 
\end{abstract}    

 \maketitle

\section{Introduction}
\begin{figure}[t]
\includegraphics[scale=0.28]{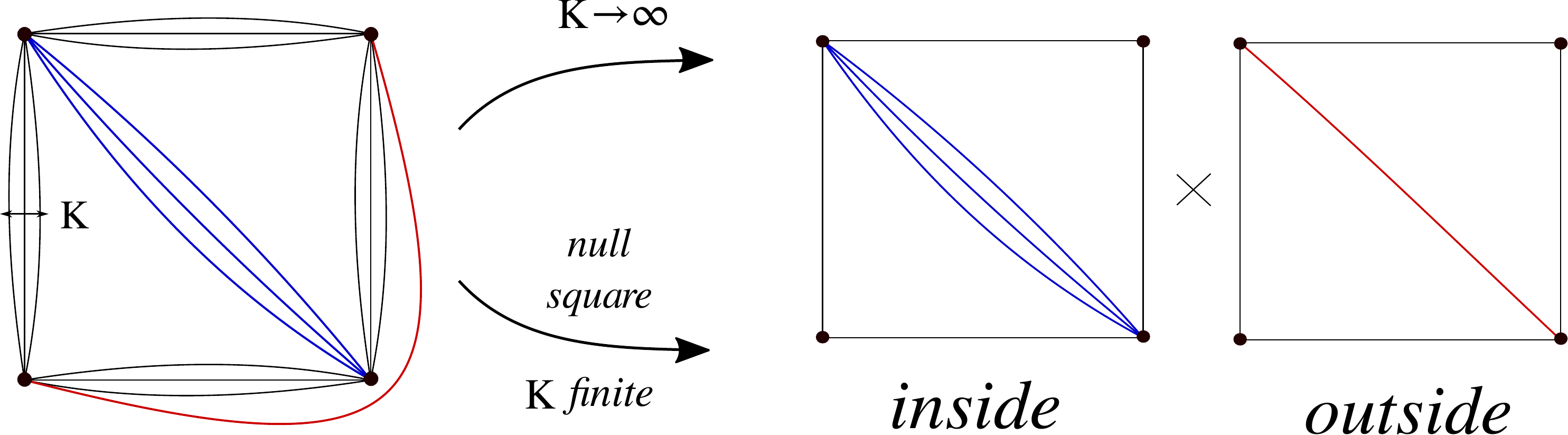}
\caption{Correlators with a large BPS frame (left) and doubly-scaled correlators with operators approaching the cusps of a null polygon (right) both factorize into products of disk topology correlators (middle).}
\label{octagonExample}
\end{figure}

Correlation functions are the backbone of quantum field theory. 
In the realm of gauge/string correspondence, correlation functions of $n$~single-trace gauge-invariant operators in the planar limit describe the scattering of $n$~closed strings, whose world-sheet is a punctured sphere. Under special circumstances this sphere simplifies further and factorizes into a product of two disks. (Reminiscent of KLT type relations connecting open and closed strings.) This can happen in at least two ways as depicted in figure \ref{octagonExample}. 

One option uses supersymmetry. In $\mathcal{N}=4$ SYM say, we can take single-trace half-BPS operators made out of \emph{many} fields and use $R$-symmetry polarizations in order to form a thick frame with many propagators sequentially connected drawing a closed polygon. The frame is SUSY protected from quantum corrections and thus the correlator splits into decoupled dynamics happening inside and outside the polygon \cite{Frank,FrankBootstrap}. 


An alternative way to decouple the inside and outside of the polygon is to make the polygon's edges approach the light-cone. If we send to zero the coupling~$g^2 $ as well as the distances between consecutive cusps~$x_{i,i+1}^2 $ with the so-called \textit{cusp times}
\begin{equation}
\label{DS}
t_k^2  \equiv \,
g^2 \log(x_{k-1,k}^2)\log(x_{k,k+1}^2)\,, 
\end{equation}
held fixed, we obtain the decoupling without any requirement on the frame's thickness \cite{stampedes1}. 
This is the regime we will focus on. This decoupling is dynamical: in this limit a fast classical particle travels around the null perimeter decoupling the quantum dynamics that are confined inside and outside the polygon \cite{Alday:2010zy,AldayBissi,BGV,BGHV,stampedes1}. The \emph{null polygon} limit (\ref{DS}) is a double-scaling limit re-summing 
the maximal power of logarithms at each perturbative order. Since it only depends
on very universal one-loop features of the conformal gauge theory, most of our results (if not all) should apply to any $4d$ conformal gauge theory with a planar limit and at least one adjoint scalar \cite{QCD}.

The key objects of this paper are these null polygons of disk topology. We think of them as a tree level skeleton which are then dressed by quantum corrections at loop order. The skeleton is described by a set of internal \textit{bridges} which are bundles of propagators connecting non-consecutive cusps. For squares and pentagons we have a single choice of such bridges (see figures~\ref{octagonExample} and~\ref{5pt_coord}) while for hexagons and higher there are multiple different topologies (see for instance figure \ref{hexagonBridges}).

Belitsky and Korchemsky found a remarkable result for the null square as  \cite{BK0,BK}  $\texttt{square}_{\ell} = e^{-s^2} \tau_{\ell}(s)$ where $\ell$ is the number of bridges, $s^2=\sum t_i^2$ and $\tau_{\ell}$ is a solution of the Toda lattice equation
\begin{equation}
\label{toda}
(s\partial_s)^2 \log \tau_{\ell}(s) = s^2  {\tau_{\ell+1}(s)\tau_{\ell-1}(s)}/{\tau_{\ell}^2(s)}\,.
\end{equation}
The square with no bridges is simply given by the Sudakov exponential $e^{-s^2}$ so that $\tau_0=1$; the exponential can be thought of as a square null Wilson loop \cite{Alday:2010zy,AldayBissi,BK0}. The square with one bridge $\tau_1$, \textit{the seed}, is a non-trivial function of $s$. Together with $\tau_0=1$ and the Toda equation it gives us recursively all other squares. Amusingly the seed itself can be fixed by the simple requirement 
%
that loop corrections in presence of $\ell$ bridges kick-in at $(\ell+1)$ loops: $\texttt{square}_{\ell} =1 + O(s^{2(\ell+1)})$ uniquely fixes $\tau_1$ to be a modified Bessel function $I_0(2s)$. All other squares then evaluate to simple determinants of Bessel functions. 

In this letter we conjecture an extension of this result and of the Toda system to any null polygon with any bridge configuration. We will encounter a novel hierarchy of Toda equations, determinantal solutions and generalized Gross-Witten-Wadia matrix integrals \cite{GWW,Brezin}. 

\section{From Stampedes to Data} 

\begin{figure}[t]
\includegraphics[scale=0.385]{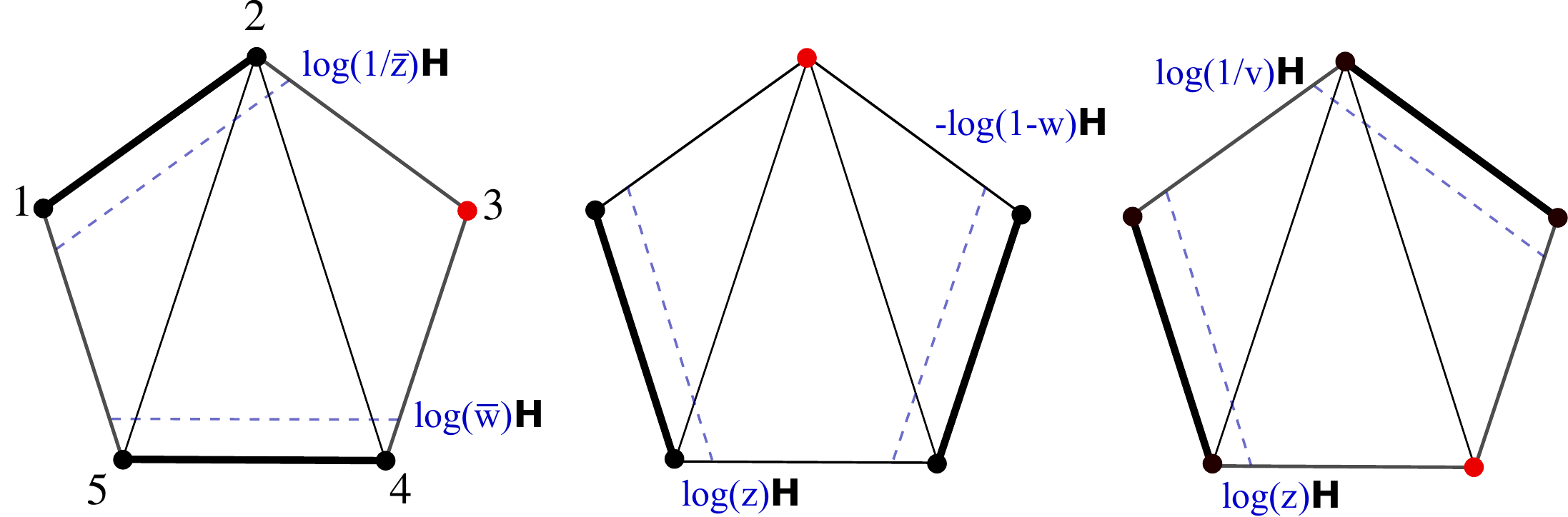}
\caption{Three of the five channels for the stampede computation of $\texttt{pentagon}$. (Reflections around the vertical axis yield the two missing channels.) Thick edges are null and expanded in $\texttt{top}$ and $\texttt{bot}$ states, then acted on by the $SL(2)$ hamiltonian $\mathbf{H}$ which captures the associated leading $\log$.  A red dot in $j$ corresponds to sub-leading cusp time $t^2_j \to 0$. } \label{pentagonChannels}
\end{figure}

We first consider pentagons. 
If we take two non-consecutive edges to approach a null limit there is a tool we can use to capture the corresponding leading logarithms: \textit{the stampedes} \cite{stampedes1}. There are five different pairs of edges we can take to be null as depicted in figure \ref{pentagonChannels}. 

\begin{figure}[t]
\includegraphics[scale=0.49]{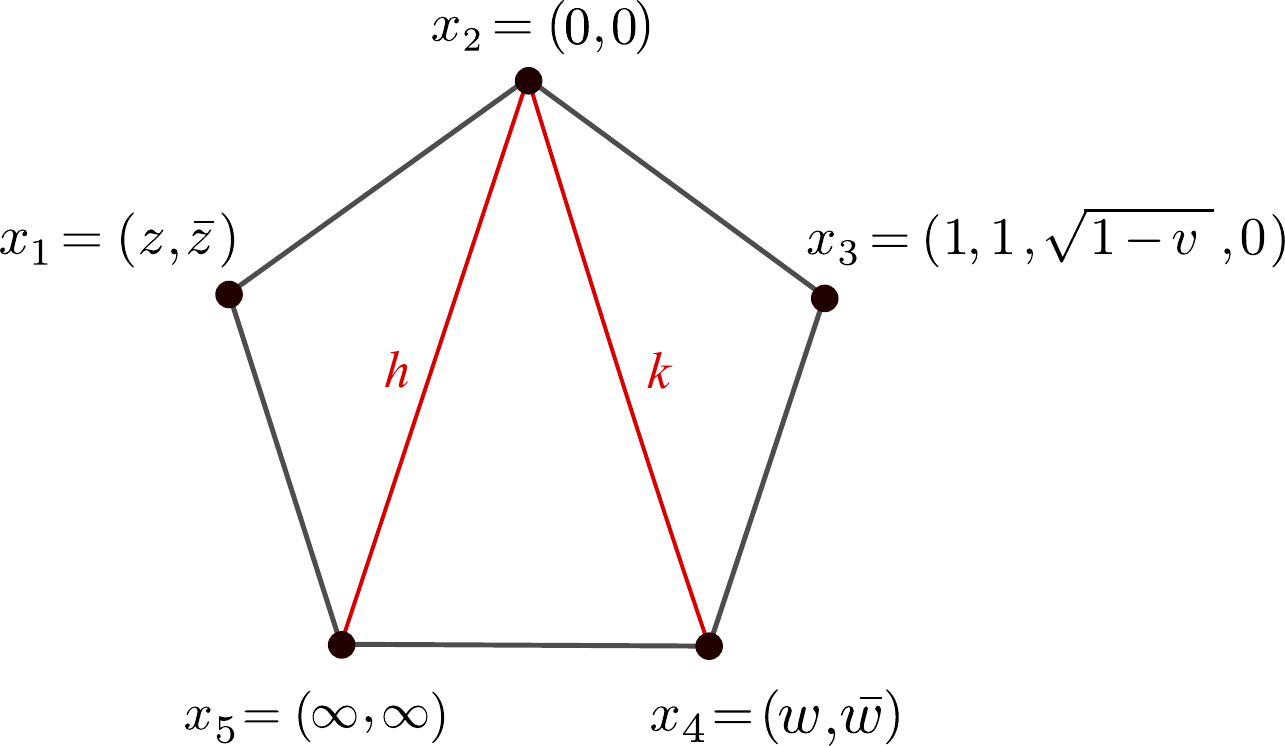}
\caption{To parametrize the pentagon we take four points on the plane as indicated in the figure and one point $
x_3$ outside the plane.
The five parameters match the five independent cross-ratios. In the null limit ${\bar z}, 1/z, 1/\bar{w}, 1-{w}, v$ go to zero.
}
\label{5pt_coord}
\end{figure}

Consider the choice $x_{12}$ and $x_{45}$ for concreteness; the other choices are treated similarly. A convenient choice of coordinates for this limit is depicted in figure \ref{5pt_coord}. In terms of these coordinates the relevant \textit{cusp times} simply read 
\begin{equation}
\begin{array}{ll}
t_1^2={\color{blue} g^2 \log(\tfrac{1}{\bar z})}{\color{magenta} \log(z)} \, ,
& t_2^2={\color{blue} g^2 \log(\tfrac{1}{\bar z})} {\color{green(pigment)} \log(\tfrac{1}{v})} \,, \\
 t_4^2 ={\color{blue}g^2 \log(\bar{w})} {\color{magenta} \log(\tfrac{1}{1-w})}  \, , & t_5^2= {\color{blue}g^2 \log(\bar{w})}  {\color{magenta} \log(z)} \,.
 \end{array} 
  \label{timesZ}
\end{equation}
The colouring is meant to highlight the different origin of these logarithms. The magenta ones arise once we Taylor expand the top and bottom edges in terms of top and bottom states
%
%
%
\begin{equation}
\label{states}
\mathcal{O}_1\mathcal{O}_2 \to \sum_{J=0}^{\infty}\frac{z^J}{J!} \,  | \texttt{top}_J \rangle\,,\,\,
\mathcal{O}_4\mathcal{O}_5 \to \sum_{J=0}^{\infty}\frac{w^{-J}}{J!} \,  \langle \texttt{bot}_J |\,,
\end{equation}
where
\begin{align*}
\begin{aligned}
| \texttt{top}_J \rangle = {  | (\partial_z^J X) \underbrace{X ... X}_{h+k+1}\rangle } \,,\,\,
 \langle \texttt{bot}_J |= { \langle  \underbrace{X... X}_{1+h} \partial_w^J(  \underbrace{X ... X}_{k+1})|}\,.
\end{aligned} \notag
\end{align*}
Here we performed already three major simplifications which we learned from \cite{stampedes1}. First, the factorization of the polygon in the limit \eqref{DS} restricts the analysis only to its interior (or, alternatively, exterior) so that $\texttt{top}$ and $\texttt{bot}$ states are mapped to vectors of an \textit{open} spin chain where each site is one field that belongs to the inside of the polygon. Second, the one-loop dilation operator that generates quantum corrections is replaced by the $SL(2)$ spin chain hamiltonian $\mathbf{H}$. That is, the leading action of the dilation on the null polygon is generated only by the \emph{hopping} of light-cone derivatives and we replaced every scalar by $X$. Finally, the thickness $K$ of the frame is irrelevant in this limit where boundary-dependant terms are sub-leading, hence we set $K=1$ \cite{generalLC}.

The blue terms in (\ref{timesZ}) appear through the action of the Hamiltonians that generate the stampedes
\begin{eqnarray}
\label{stampede}
&&\texttt{stampede}^{h,k}_{J,J'}=  \\
&&\qquad =\langle \texttt{bot}_{J'} |   e^{g^2 \log(\bar{w})  \mathbf{H} } \big(\mathbb{I} \otimes |v\rangle \langle v |  \big) e^{g^2 \log(1/\bar{z})  \mathbf{H}} | \texttt{top}_{J} \rangle\,.  \nonumber
\end{eqnarray}
Dependence on $h,k$ is implicit on the rhs. The stampede is given by bottom and top time evolution together with the overlap with the impurity at the right-most site which is the sole effect of the last operator, the off the plane one. For that we use the simple scalar product 
\begin{equation*}
\langle J |v\rangle \langle v | J' \rangle  = \frac{1}{v^{J}}  \times 1 \,, 
\end{equation*}
see appendix B for more details. In the end we have 
%
\begin{equation}
\label{F_zwv}
\mathcal{F}_{h,k}\equiv \sum_{J,J'=0} \frac{z^J}{w^{J'}} \texttt{stampede}^{h,k}_{J,J'}\,. 
\end{equation}
For example for two equal bridges of length $1$ the pentagon starts at two loops and evaluating the smallest stampedes for the lowest spins $J,J'\leq 1$ yields
\begin{eqnarray}
\mathcal{F}_{1,1} &=& 1+ \frac{z}{v}(-g^6 \log^3 \bar z+\dots)+ \nonumber \\
&&\!\!\! + \frac{z}{w}(g^4 \log^2 \bar{z}/\bar w +4 g^6 \log^3 \bar{z}/\bar w +g^6 \log^3 \bar{z} +\dots) \nonumber \\
&&\!\!\!+\frac{1}{w} (-g^4 \log^2 \bar{w}+ 4 g^6 \log^3\bar w+\dots)+\dots \,.
 \label{expansion}
\end{eqnarray} 
Such expansions are very easy to generate to large orders in $z$ and $1/w$ and $1/v$ (i.e. large spins) and to relatively large order in $g^2$ (we have gone to five loops, i.e.~$g^{10}$). They are what we call our \textit{data}.

\begin{figure}[t]
\includegraphics[scale=0.52]{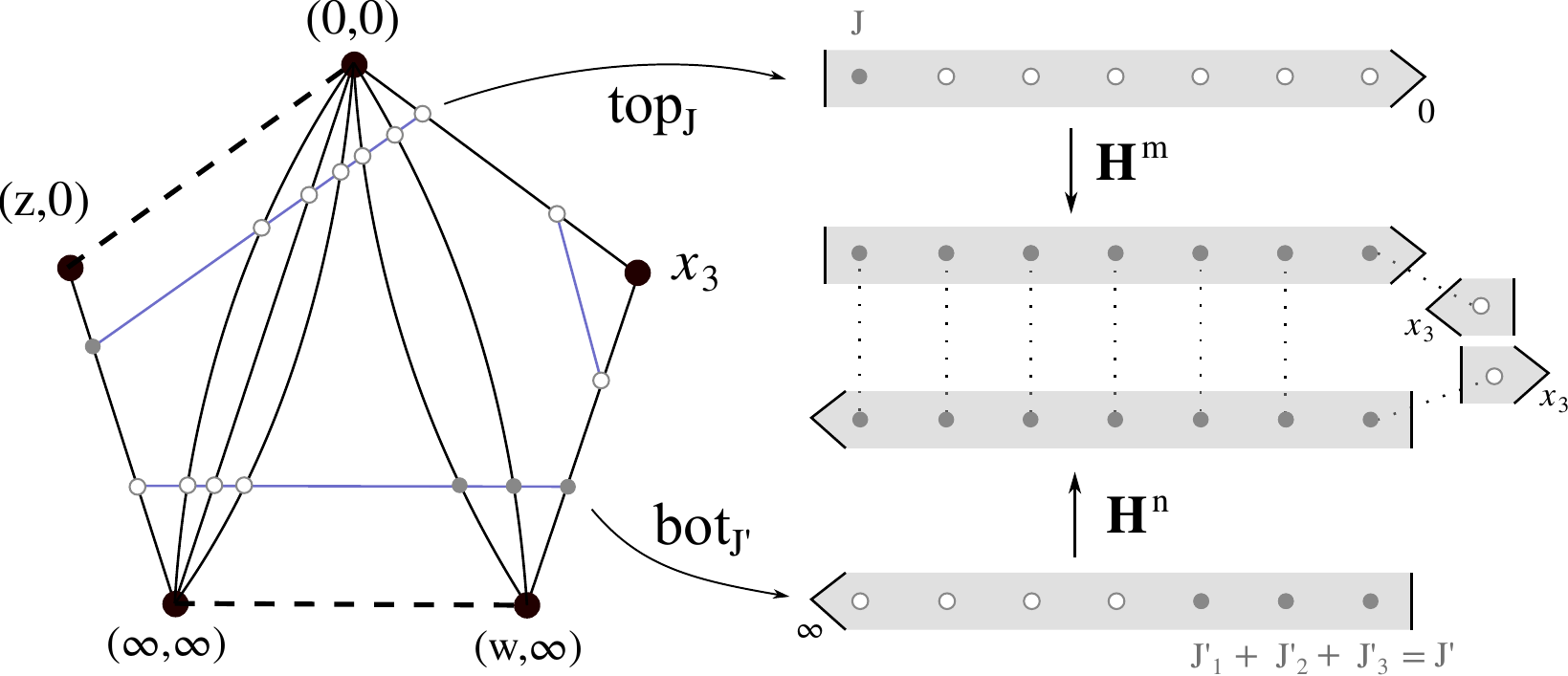}
\caption{The Taylor expansion of null edges (dashed) at order $J/J'$ around $0/\infty$ creates $\texttt{top}_{J}/\texttt{bot}_{J'}$ states, depicted here for $h=3$, $k=2$. Each chain site is a circle: filled ones are those where derivatives are injected. The states $\texttt{bot}$ and $\texttt{top}$ evolve by the $SL(2)$ action at order $(g^2 \log 1/\bar z)^m$ and $(g^2 \log\bar w)^n$. The resulting states can have derivatives distributed  among all sites. Overlap site-by-site is illustrated by dotted lines.}
\label{fig:np_fact}
\end{figure}


%
%
%
%
%

\section{From Data to Symbols to Pentagons} 
We now need to re-sum (\ref{expansion}) which is an expansion with small $z, 1/v, 1/w$ while we are after the null limit where these three variables are either finite or very large.
This is easy. We found that at $L$ loops we can express the full functions of these three variables in terms of 
iterated Goncharov polylogarithms of trancendentality $L$ which can be nicely coded in terms of symbols \cite{symbols}. The symbol ansatz is generated as a linear combination of \emph{words} $a_1 \otimes a_2 \otimes \cdots \otimes a_L$ of \emph{letters} $a_i$ picked from the set
\begin{equation}
\label{symbolSet}
\{z,w,v,1-z,1-w,1-v,w-z,v-z,v- w\}\,.
\end{equation}
(more in appendix C.)
At two loops, for instance, the expansion (\ref{expansion}) is nothing but the Taylor expansion of 
\begin{equation}
\notag
 \log^2 \!\bar w\left[\text{Li}_2({1}/{w}) +\tfrac{1}{2} \log ^2\!\left(1\!-\!{1}/{w}\right)\right]\!-\! \log^2\! ({\bar w}/{\bar z} )\text{Li}_2(z/w)\,,
\end{equation}
whose symbol is
\begin{equation}
\la{symbol11}
\!\!  \log^2 \!\bar w\left[  \frac{1}{1-w} \otimes \frac{1-w}{w}\right] \! + \log^2\! ({\bar w}/{\bar z} )\left[ \frac{z-w}{w} \otimes \frac{z}{w} \right]\,.
\end{equation}
Stampedes count movements of bosons (the derivatives) in a spin chain leading to nice rational numbers with combinatorial interpretations. There are no $\pi$'s or $\zeta$'s in expansions like (\ref{F_zwv}) so the symbol misses no information in this case! We simply write an ansatz as in (\ref{symbol11}) with unknown rational coefficients multiplying each word which we then fix by matching with expansions such as \eqref{F_zwv}.

We can then expand the resulting functions in the null limits of interest. For example 
\begin{equation*}
\lim_{z\to \infty} \lim_{w \to 1} \lim_{v\to 0}\mathcal{F}_{h,k} = \texttt{pentagon}_{h,k}(t_1,t_2,0,t_4,t_5)\,.
\end{equation*}
Applied to the example $h=k=1$ we get 
\begin{align*}
\begin{aligned}
&\left.\texttt{pentagon}_{1,1}\right|_{t_3=0} =1- \tfrac{1}{4} \left(2\, t_1^2
   t_5^2+t_1^4+t_4^4+t_5^4\right)\\&\qquad\qquad\qquad + \tfrac{1}{36} (4\, t_4^6 - 3\, t_1^2 t_2^4 - t_2^6) +\tfrac{1}{9} (t_1^2 + t_5^2)^3+\dots
\end{aligned}
\end{align*}
Considering different channels each time, we can generate a huge amount of null \texttt{pentagons} with a single cusp time set to zero. This allowed us to fix all \texttt{pentagons} up to a \textit{single} constant at five loops when the first completely mixed term $t_1^2t_2^2t_3^2t_4^2t_5^2$ appears. This term can never be detected if we send any of the five cusp times to zero. The $\texttt{pentagon}_{1,1}$ for example is given in appendix D up to five loops.

\section{A gift from the stampedes} 
How can we use all these \texttt{pentagons} computed to high perturbative order 
to unveil an underlying structure leading to their exact expressions? 
We looked for a set of differential equations for these objects akin to the Toda equation (\ref{toda}) for the square. We made an educated ansatz for such equations and used the stampede data to fix their form. This leads to the main formulae of this letter: 
%
%
%
%
\begin{align}
\begin{aligned}
\label{Todas}
&\left(\mathcal D_1+\mathcal D_2\right)\left(\mathcal D_1+\mathcal D_5\right) \log \mathbb{P}_{h,k}= 4t_1^2 \, \frac{\mathbb{P}_{h+1,k} \mathbb{P}_{h-1,k}}{\mathbb{P}_{h,k}^2}\,,\\
&\left(\mathcal D_3+\mathcal D_2\right)\left(\mathcal D_3+ \mathcal D_4\right) \log \mathbb{P}_{h,k}= 4t_3^2\,  \frac{\mathbb{P}_{h,k+1} \mathbb{P}_{h,k-1}}{\mathbb{P}_{h,k}^2}\,,
\end{aligned}
\end{align}
where $\mathcal{D}_j=t_j \partial/\partial t_j$ and 
\begin{equation}
\texttt{pentagon}_{h,k}=e^{-t_1^2-\dots-t_5^2}\times \mathbb{P}_{h,k}(t_j)\,,\,\,\,\, \mathbb{P}_{0,0}=1\,.
\end{equation}

 A second crucial piece of information from the stampede are the reductions of $\texttt{pentagon}_{h,k}$ when only specific cusp times are switched on. In particular, when the times $t_1,t_3$ are set to zero equations \eqref{Todas} imply factorization. Setting most t's to zero degenerates the pentagon into squares so we expect these functions to be given by squares. We find precisely such square reductions as
 \begin{equation}
 \label{pentared}
 \mathbb{P}_{h,k}(0,t_2,0,t_4,t_5)= {\tau}_{h}(t_5) {\tau}_{h+k}(t_2) {\tau}_{k}(t_4)\,,
\end{equation}
where each $\tau$-function has an index given by the total number of bridges emitted by the $i$-th cusp.

Henceforth we take the stampedes gift (\ref{Todas},\ref{pentared}) as our starting point and explain how these equations fix \emph{all} polygons at \emph{any} loop order.
\section{Toda Bootstrap}
Under the assumption that $\mathbb{P}_{h,k}$ has a regular Taylor expansion around $t_j ^2=0$
\begin{equation}
\mathbb{P}_{h,k} =\sum_{n_1\dots n_5} \mathbbm{c}_{h,k}(n_1,\dots,n_5) \times  t_1^{2n_1}\cdots t_5^{2n_5}\,, \la{startBootstrap}
\end{equation}
the first of \eqref{Todas} can be used to express any coefficient $\mathbbm{c}(n_1,\dots,n_5)$ as a function of coefficients $\mathbbm{c}_{h,k},\mathbbm{c}_{h\pm 1,k}$ with lower value of $n_1$, due to the factor $4t_1^2$ in the r.h.s. of the equation. Likewise, the second equation in \eqref{Todas} can be used to express each $\mathbbm{c}_{h,k}$ via coefficients with lower $n_3$. The iterative application of the two equations eventually reduces the problem to the determination of
\begin{equation}
\notag
\mathbbm{c}_{h,k}(0,n_2,0,n_4,n_5)\,,\,\,\, \forall \,h,k \in \mathbb{N}\,.
\end{equation}
But these coefficients are precisely those determined by the reduction (\ref{pentared}) so we are done! 

Pertinent for higher polygon generalizations, an important observation 
 is that $\mathbb{P}_{h,0}$ only only depends on times which are either end-points of the bridges $(t_2,t_5)$ or sandwiched between two such points $(t_1)$; the two times $t_3$ and $t_4$ do not show up. For the first such pentagon we managed to identify its exact re-summation as $\mathbb{P}_{1,0} = \phi(t_1,t_2,t_5)$ and $\mathbb{P}_{0,1} = \phi(t_3,t_2,t_4)$, where
\begin{equation}
\phi(x,y,z) = \sum^{\infty}_{n=0}\sum_{m}^n \sum_{r}^{n-m} \frac{x^{2(n-m-r)} y^{2 m}z^{2 r}}{(n-m)!(n-r)!m!r!}\,. \label{phi}
\end{equation}
Any solution $\mathbb{P}_{h,k}$ with $h,k\geq 1$ can be cast in the compact form of a nested determinant
\begin{align}
\begin{aligned}
\label{determinant}
&\mathbb{P}_{h,k}\! =\! \frac{\begin{vmatrix}\left(\mathcal D_1\!+\!\mathcal D_2\right)^{i-1}\left(\mathcal D_1\!+\!\mathcal D_5\right)^{j-1} \mathbb{P}_{1,k}
\end{vmatrix}_{i,j\leq h}}{(2t_1)^{h-h^2} \,\mathbb{P}_{0,k}^{h-1}} \,,\\
&\mathbb{P}_{1,k}\! =\! \frac{ \begin{vmatrix} \left(\mathcal D_3\!+\!\mathcal D_2\right)^{i-1}\left(\mathcal D_3\!+\!\mathcal D_4\right)^{j-1} \mathbb{P}_{1,1}
\end{vmatrix}_{i,j\leq k}}{
(2t_3)^{k-k^2} \mathbb{P}_{1,0}^{k-1}   }
\,.
\end{aligned}
\end{align}

\begin{figure}[t]
\includegraphics[scale=0.235]{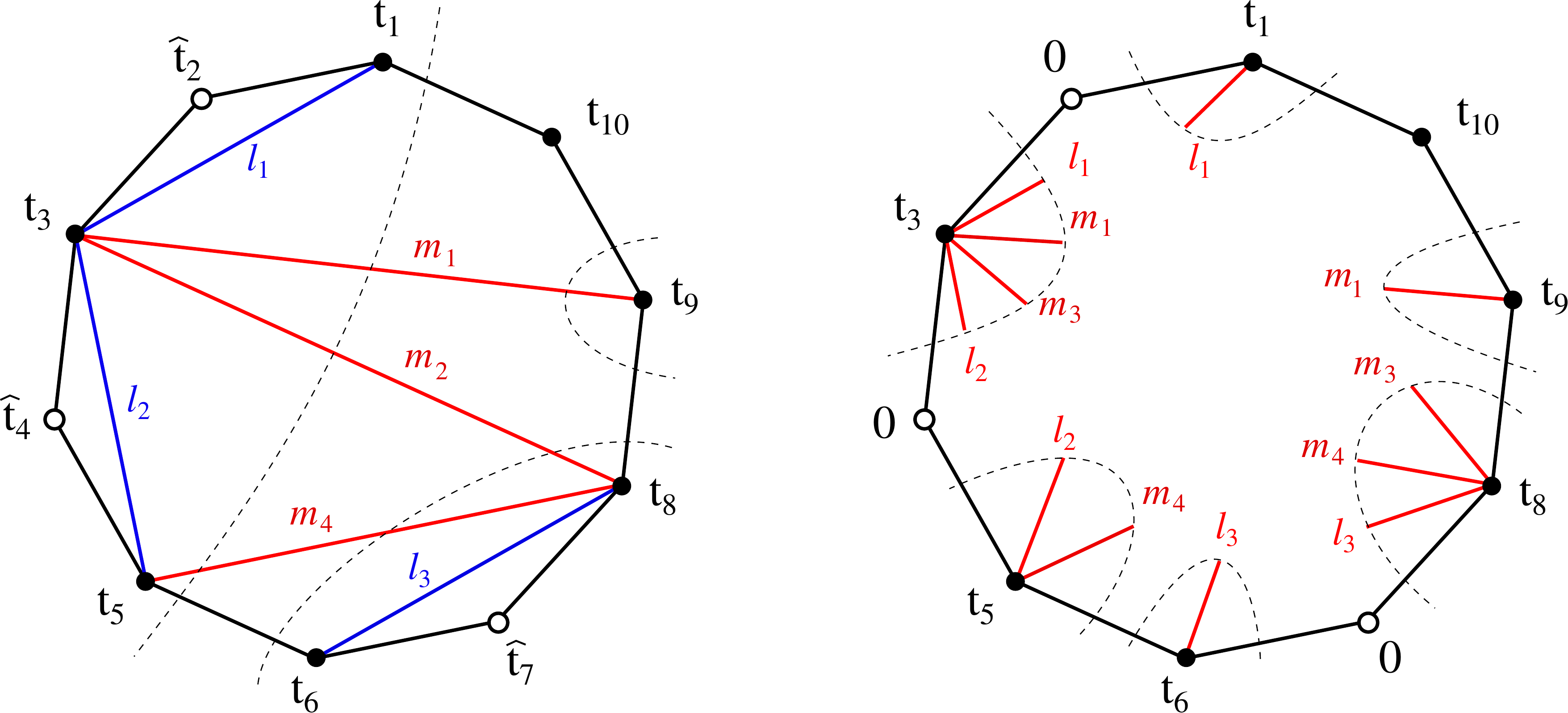}
\caption{A decagon example with blue bridges, stretching between next-to-neighbouring points, and red bridges. Dashed lines cut the null polygon into factorized contributions, each subject to a set of coupled Toda equations. On the right we depicted the boundary data at $t_j = 0$ which fixes the solution.}
\label{fig:factor_n}
\end{figure}
This representation follows iteratively from \eqref{Todas} and from the fact that $\mathbb{P}_{0,k}$ and $\mathbb{P}_{h,0}$ are functions of three cusp times only. Formulae \eqref{determinant} make manifest that everything can be reduced to \emph{seed} functions $\mathbb{P}_{0,1}$ and $\mathbb{P}_{1,1}$. The former is given by (\ref{phi}) while the latter can be bootstrapped by the self-consistency argument presented above; we present it in appendix D up to five loops.  
\section{Higher polygons}
Based on the experience collected on \texttt{pentagon} functions, we can readily conjecture the form of any null polygon. We express a polygon with $n$ edges as the product of an exponential of cusp times -- reminiscent of the Sudakov term in a null polygonal Wilson loop -- and a function $\mathbb H$ that depends on the configuration of bridges
\begin{equation}
\texttt{polygon} = e^{-\sum_i t_i^2} \times \mathbb{H}(t_1,\dots,t_n)\,.
\end{equation}
When there are no bridges, as for instance for correlators of short operators, we known from \cite{BGV} that $\mathbb{H}$ is one.

\begin{figure}[t]
\includegraphics[scale=0.36]{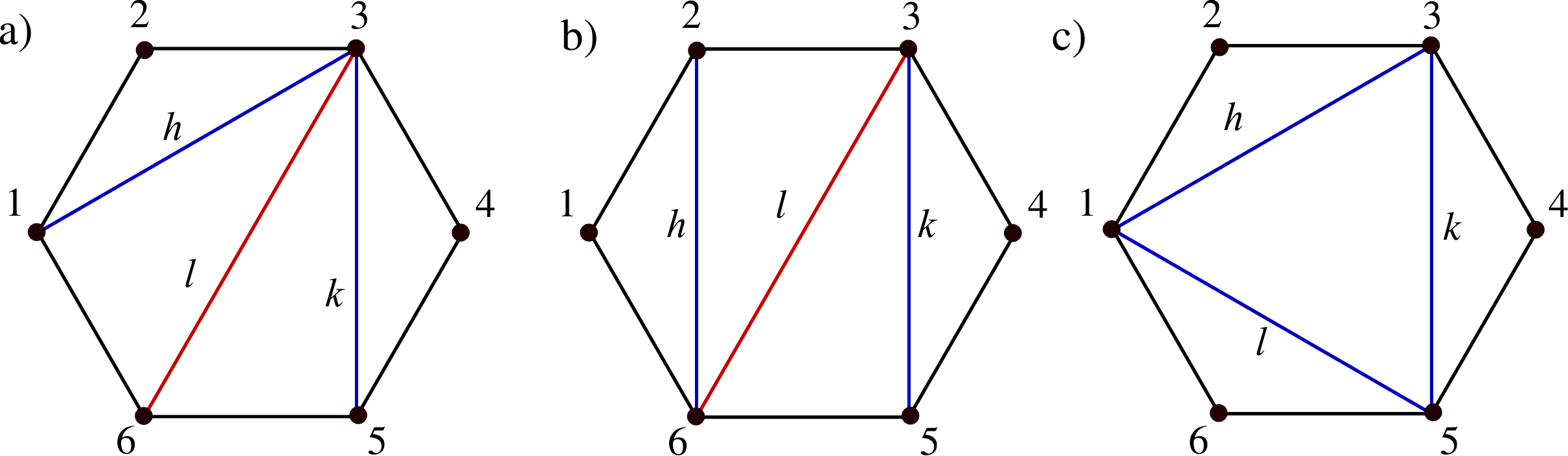}
\caption{The three possible configurations of bridges for the $\texttt{hexagon}$ in the planar limit. Blue bridges are associated to $2d$ Toda equations of type \eqref{Todas}.}
\label{hexagonBridges}
\end{figure}

Let $\ell$ be the number of bridges stretching between $t_{i-1},t_{i+1}$, we denote $\hat t_i$ the cusp time comprised between the end-points. We focus on this bridge and write $\mathbb{H}_{\ell}$ leaving other bridge indices implicit. We can state that:
\begin{itemize}
\item there is a Toda equation of type \eqref{Todas} in $t_{i-1},\hat t_i,t_{i+1}$ relating $\mathbb{H}_{\ell}$ to $\mathbb{H}_{\ell \pm 1}$. When $\hat t_i=0$ it becomes a factorization condition $\mathcal{D}_{i-1} \mathcal{D}_{i+1} \mathbb{H}_{\ell}=0$. 
\item the function $\mathbb{H}$ depends only on cusp times that emit bridges (e.g. $t_{i-1},t_{i+1}$) or are sandwiched between the two endpoints of a bridge (e.g. $\hat t_{i}$).
\item when $\hat t_{j}=0$, the function $\mathbb{H}$ factorizes into a product of squares $\tau_{n}(t_k)$, one for each cusp $k$ emitting a total of $n$ bridges.
\end{itemize}
This set of statements traces back to $\texttt{pentagon}$ observations. For higher polygons the novel feature is the presence of both blue bridges in figure \ref{fig:factor_n} -- connecting next-to-nearest neighbouring cusps $t_{i\pm1}$ and governed by a $2d$ Toda equation -- and the red bridges which connect further apart cusps. To complete the conjecture we add:
\begin{itemize}
\item The polygon $\mathbb{H}$ is factorized into the product of functions $\mathbb{X}$ of those cusps $t_i$ and $\hat t_i$ comprised between the end-points of a continuous path of blue bridges. (In figure \ref{fig:factor_n}, for instance, we have three such blue paths of lengths $2$, $1$ and $0$ and thus the correlator would factor into three factors.) Paths of length $0$ correspond to $\texttt{square}_n(t_j)$ where n is the total number of bridges emitted by the cusp $t_j$. 
\end{itemize} 
Let us apply these conjectures to the general \texttt{hexagon}. A planar hexagon can have three different bridge configurations as depicted in figure \ref{hexagonBridges}. Cases $(a)$ and $(b)$ in the figure feature two blue bridges, connected and disconnected respectively. Hence they are subject to two $2d$ Toda equations as does \texttt{pentagon} to which they are closely related.
Case $(a)$ is factorized as
\begin{equation}
\mathbb{H}^{(a)}_{h,l,k} =  \mathbb{X}_{h,l,k}(t_1,\hat t_2,t_3, \hat t_4,t_5) \tau_{l}(t_6)\,.\notag
\end{equation}
The function $\mathbb{X}$ solves two coupled equations in $(t_1,\hat t_2,t_3)$ and $(t_3,\hat t_4,t_5)$ moving $h$ and $k$. (Notice that for $l=0$ the same holds for $\texttt{pentagon}$, which means $\mathbb{X}_{h,0,k}=\mathbb{P}_{h,k}$.) Case $(b)$ features two disconnected blue bridges, i.e. it solves decoupled equations and is thus 
factorized into two functions of three cusps and expressed via $\texttt{pentagon}$ functions already determined in this letter,
\begin{equation}
\mathbb{H}^{(b)}_{h,l,k}=\mathbb{P}_{l,h}(0,t_6,\hat t_1,t_2,0) \mathbb{P}_{l,k}(0,t_3,\hat t_4,t_5,0)\,.\notag
\end{equation}
Case $(c)$ features three connected blue bridges
and thus is solution to three pair-wise coupled $2d$ Toda equations in $(t_1,\hat t_2,t_3)$, $(t_3,\hat t_4,t_5)$ and $(t_5,\hat t_6,t_1)$ moving $h$, $k$ and $l$ respectively. The boundary data for the three cases are
\begin{eqnarray*}
\mathbb{H}^{(a)}_{h,k,l}(t_1,0,t_3,0,t_5,t_6)&=& \tau_{h}(t_1) \tau_{h+k+l}(t_3) \tau_{k}(t_5) \tau_l(t_6) \,, \nonumber\\
\mathbb{H}^{(b)}_{h,k,l}(0,t_2,t_3,0,t_5,t_6)&=&\tau_{h}(t_2) \tau_{h+l}(t_6) \tau_{k+l}(t_3)  \tau_{k}(t_5)\,,\nonumber \\ 
\mathbb{H}^{(c)}_{h,k,l}(t_1,0,t_3,0,t_5,0)&=& \tau_{h+l}(t_1) \tau_{h+k}(t_3) \tau_{k+l}(t_5)\,.
\end{eqnarray*}
We tested these various statements against stampede data at three loops for a few six-point correlators \cite{hexa_hop}.

\section{Discussion}

In this paper we used the stampede technology developed in \cite{stampedes1} to unveil new integrable structures in the null limits of conformal gauge theories. The final picture is quite compact. In the double-scaling limit~\eqref{DS}, correlators are characterized by internal bridges which can be blue (if they connect next-to-neighbouring cusps) or red (if they connect farther separated vertices). With zero blue bridges the correlators factorize into products of square functions (the boundary data); finite blue lengths can then be constructed recursively by means of Toda equations \eqref{Todas} \cite{GrishaHint}.

Some of the solutions to these PDEs are rather novel. If we take a five point correlator with two non-zero cusp times, for instance, we find 
\begin{align*}
\begin{aligned}
&\mathbb{P}_{h,k}(t_1,t_2,0,0,0) = \det_{1\leq i,j\leq {h+k}} \begin{vmatrix} z_i^{i-j} I_{i-j}\left(2 z_i \right)\end{vmatrix}\,,
\end{aligned}
\end{align*}
where $I_j$ are modified Bessel functions of the first kind and $z_i=\sqrt{t_1^2+t_2^2}$ if $i\leq h$ and $z_i=t_2$ for $i>h$. It admits a 
representation as an exotic matrix integral
\begin{equation}
\notag
 \int \! \frac{d \theta_1}{2\pi}\cdots\!\int\! \frac{d\theta_{\ell}}{2\pi} \Delta(Z) e^{\text{Tr}(Z+Z^{\dagger})} \prod_{k} (z_k e^{i \theta_k})^{1-k},
 \end{equation}
where $Z$ is a matrix with eigenvalues $\{z_j e^{i\theta_j}\}$ and $\Delta$ is the Vandermonde determinant. (Details in appendix E.) When $x_i=y$ this integral reduces to the Gross-Witten-Wadia matrix model \cite{GWW}. What is the physical origin of these matrix integrals? 

It would be fascinating to connect the structures found here to other exact approaches to quantum gauge theories. We can envisage three natural connections.

A first connection would be with conformal bootstrap explorations. In \cite{AldayBissi,BGV,BGHV,Antunes} the conformal bootstrap was applied to the study of null correlation functions. There, no simplifying double-scaling limit was taken. On the other hand the analysis was restricted to correlation functions of the lightest fields of the theory which translates into polygons with no internal bridges.
Indeed, if we take the Sudakov and recoil factors bootstrapped there, in the limit (\ref{DS}) we would obtain the universal exponentials 
\begin{equation}
e^{-t_1^2-\dots - t_n^2}\,
\end{equation}
 showing up everywhere in our letter. When the bridges are non-zero, dressing these exponentials we have the very nontrivial solutions to the Toda hierarchy. Would be remarkable to fit them inside a bootstrap analysis. Combining the technology developed here with a bootstrap approach might lead to a more general null limit for any polygon. The physical picture should be one where in the null limit correlators become Wilson loops dressed by further cusp insertions of adjoint fields. The latter then propagate from cusp to cusp interacting with the planar chromodynamic flux tube of the theory.
 
 We computed the hexagon for any bridge topology as depicted in figure \ref{hexagonBridges}; from an OPE perspective the last one resembles a snowflake while the other two mimic the structure of comb channel decompositions. Despite some encouraging first steps \cite{Goncalves:2019znr, Rosenhaus:2018zqn}, there is no analytic result for comb channels for $n > 4$ points. Can one use our hexagon results to improve this state of affairs?
   
A second connection would be with integrability. In planar $\mathcal{N} = 4$ SYM four-point disk correlators were computed for any kinematics in \cite{Frank,FrankBootstrap} by means of bootstrap plus hexagonalization \cite{hexagons,Hexagonalization,morehexagons2,Determinant} and preliminary higher polygons explorations were started in \cite{decagons,Hexagonalization2}. Would be very interesting to  connect our results with these. In the hexagonalization formalism, for instance, the limit~\eqref{DS} is governed by an interesting region where all rapidities and bound-state numbers of the mirror particles are (analytically continued to) zero. The square was identified with an amplitude in the Coulomb branch \cite{FrankSimon}; are the doubly-scaled higher polygons found here higher-point scattering amplitudes in some interesting kinematics?

A last connection could be with supersymmetric localization. Determinants of Bessel functions and Toda equations  constantly pop up in such supersymmetric studies, see e.g. \cite{GiombiShota,KolyaBremss,JaumeZohar} --  is this a coincidence?

Finally, can we fix the so called pentagon at all loops following a bootstrap \`{a}-la \cite{Frank,stampedes1} for the square (called octagon there)? A strategy would be to construct a symbol ansatz for five point functions of large operators. (We already know some of its letters from (\ref{symbolSet}) arising from the stampede analysis here.) Full null polygon limits will not be enough to fix the ansatz but single light-cone limits – which are still fully captured by the stampedes – might suffice once combined with simple additional physical conditions such as Steinmann relations and single-valuedness in the physical sheet. 

\begin{acknowledgments}
\section*{Acknowledgments}
\label{sec:acknowledgments}

We thank B.~Basso, A.V.~Belitski,  C.~Bercini, V.~Goncalves, A.~Homrich, S.~Komatsu and especially G.P.~Korchemsky for discussions and valuable comments on our draft.
Research at the Perimeter Institute is supported in part by the Government of Canada 
through NSERC and by the Province of Ontario through MRI. This work was additionally 
supported by a grant from the Simons Foundation (Simons Collaboration on the Nonperturbative Bootstrap \#488661) and ICTP-SAIFR FAPESP grant 2016/01343-7 and FAPESP grant 2017/03303-1. \end{acknowledgments}

\appendix

\section{A. Explicit Polarized Correlators}
We can consider a five-point function in the planar limit of the $\mathcal{N}=4$ SYM gauge theory
\begin{equation}
\notag
 \langle \mathcal{O}_1(x_1) \mathcal{O}_2(x_2) \mathcal{O}_3(x_3) \mathcal{O}_4(x_4) \mathcal{O}_5(x_5) \rangle\,,
 \end{equation}
and choose the half-BPS operators to be
\begin{eqnarray}
\nonumber
\mathcal{O}_1&&= \text{Tr}\,\bar{X}^{K}Z^{K}\,,\,\mathcal{O}_2= \text{Tr}\,X^{2K}Y^{\ell +m}\,,\,\mathcal{O}_3= \text{Tr}\,\bar X^{K} Y^{K}\,,\\
\mathcal{O}_4&&= \text{Tr}\,\bar Z^{2K} \bar Y^{\ell}\,,\,\mathcal{O}_5= \text{Tr}\,Z^K \bar Y^{K+m}\,,\label{operators}
\end{eqnarray}
where $\text{Tr}$ is the trace over $SU(N)$ indices of fields plus their permutations. 
By means of a conformal transformation four points are sent to a plane, described by coordinates $x_{\pm} = x^0 \pm x^1$, as in figure \ref{5pt_coord}.
The fifth point lies outside of the plane at $x^{\mu}_3=(1,1,\sqrt{1-v},0)$ (the first two components are $x_{\pm}$; the last two are cartesian). The variables $(z,\bar z,w,\bar w,v)$ parametrize the five independent cross-ratios
\begin{equation}
U_k=\frac{x_{k,k+1}^2x_{k+2,k-1}^2}{x_{k,k+2}^2x_{k-1,k+1}^2}\,,
\end{equation}
for $k=1,\dots,5$ and with the identification $x_{k}=x_{k + 5}$\,.
In this coordinate frame the cross-ratios read
\begin{align*}
\begin{aligned}
&U_1=\frac{z\bar z}{v-z-\bar z +z\bar z}\,,\, U_2= \frac{v(z-w)(\bar z - \bar w)}{(v-z -\bar z + z \bar z) w\bar w}\,,\\
&U_3 =\frac{v-w- \bar w+w \bar w}{w \bar w} \,,\, U_4 =\frac{v-z -\bar z + z \bar z}{(w-z)(\bar w-\bar z)} \,,\\&U_5 =\frac{w \bar w}{(w-z)(\bar w-\bar z)}\,,
\end{aligned}
\end{align*}
hence the null polygon limit $U_k\to 0$ can be reached as 
\begin{equation}
z\to \infty\,,\, \bar z \to 0\,,\, w\to1\,,\,\bar w\to \infty\,,\, v\to 0\,.
\end{equation}
\begin{figure}[t]
\includegraphics[scale=0.6]{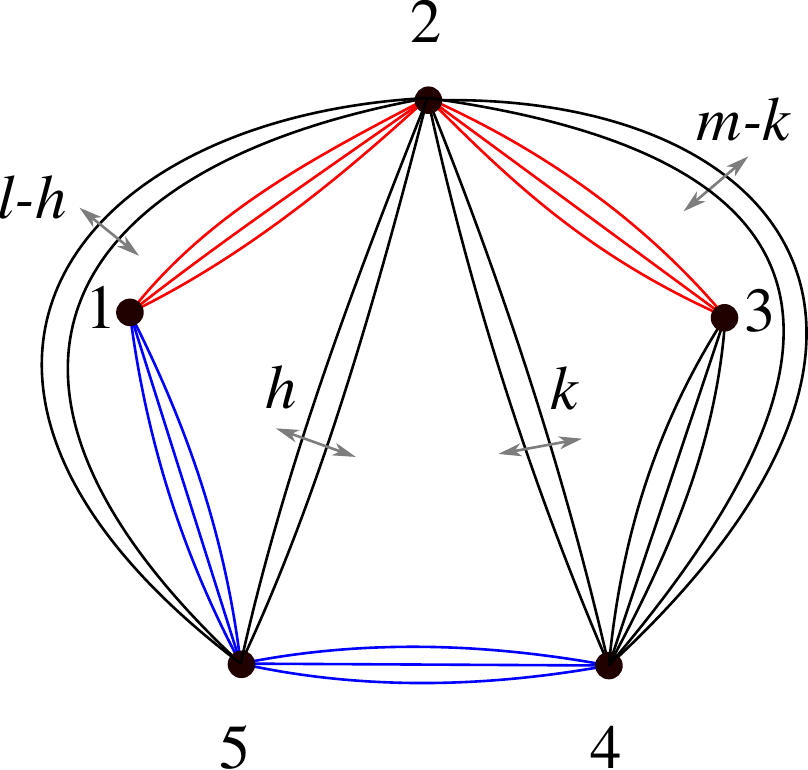}
\caption{Five-point skeleton graph with $(h,k)$ internal bridges. Red, blue, black lines denote propagators of fields $X,Z,Y$.}
\label{skeleton_appendix}
\end{figure}
In the planar limit the only tree-level graphs contributing to the five-point correlator are given by a frame of exactly $K$ propagators connecting consecutive operators, plus a number $(l,m)$ of propagators passing inside or outside the pentagonal perimeter, as depicted in figure \ref{skeleton_appendix}. For the choice of polarizations \eqref{operators}, different graphs are simply distinguished by the number $(h,k)$ of internal bridges versus $(l-h,m-k)$ external ones.

\section{B. Spin Chain}
The three simplifications alluded to below \eqref{states} can easily be relaxed to describe a much more general limit where only $\bar w$ and $1/\bar z$ are sent to zero with the other three cross-ratios kept fixed. We call it the single light-cone limit. It can still be described by the stampedes but we would need to refine the starting point (\ref{states}) replacing it by
\begin{align}
\begin{aligned}
\label{statesGen}
&|\texttt{top}_J\rangle = |(\partial^J Z^K) (X^{K}Y^{l+m}+\dots)\rangle_{\texttt{closed}}\,,\\
&\langle \texttt{bot}_J|  =  \langle (Y^{l-h} Z^{K}Y^{h} +\dots)\partial^J (Y^{m+K})|_{\texttt{closed}}\,,\\
&|v\rangle = |X^k\rangle_{\texttt{closed}} \,,\,\,\, \langle v|=\langle Y^k |_{\texttt{closed}}\,,
\end{aligned}
\end{align}
where $\dots$ stand for the sum over permutations, reminiscent of the half-BPS definition.
We would then need to act on the bottom with the full $PSU(2,2|4)$ spin chain Hamiltonian with its full harmonic action representation, see \cite{Beisert,SVW,stampedes1}. 
In contrast in the main text due to the simplification of replacing all scalars by a single scalar $X$ we can simply use the much simpler $SL(2)$ spin chain Hamiltonian ${\bf H}= \sum_{n} {\bf h}_{n,n+1}$ with 
\begin{equation*}
\textbf{h} |n,m\rangle\! = \!-(\mathbbm{h}(n)+\mathbbm h(m)) |n,m\rangle+\!\! \!\sum_{j\geq0,j\neq n}^{m+n} \! \frac{ |j,m+n-j\rangle}{|n-j|}\,,
\end{equation*}
where $\mathbbm h(n)$ are Harmonic numbers and
\begin{equation}
\notag
|J\rangle = \frac{1}{J!} \partial^J X,\,\,\, \partial = \frac{1}{2}\left(\partial_0 + \partial_1\right)\,.
\end{equation}
Scalar product are derived from the bare propagator $
{}_x\langle 0|0\rangle_{y} = {1}/{(x-y)^2}\,,$
by acting with derivatives and conformal transformations on the points $x,y$.
Relevant scalar product for the stampede are
\begin{align}
\begin{aligned}
\label{sc_prods}
&{}_{\infty}\langle J|J'\rangle_0 = \delta_{J,J'}\,,\,\,\,\,{}_{x}\langle 0|J\rangle_0 ={x_-^J}/{(x^2)^J}
\,,\\
&{}_{\infty}\langle J|0\rangle_y = {}_{\infty}\langle J|e^{y_+ \partial}|0\rangle_0 = \sum_{J'}y_+^{J'}  {}_{\infty}\langle J|J'\rangle_0 = y_+^J\,.
\end{aligned}
 \end{align}
From \eqref{sc_prods} it follows that
\begin{equation}
{}_{\infty}\langle J'|v \rangle \langle v|J \rangle_0  = \frac{1^{J+J'} }{(1+v-1)^J}\,.
\end{equation}
In order to match data with an ansatz, the stampede function should be normalized by its tree level value,
which is not just $1$ for $\texttt{pentagon}$ and any higher $\texttt{polygon}$ even after sending one point at infinity and setting the expanded edges on the light-cone. The \texttt{pentagon}, for instance,  we read it out from formula \eqref{stampede} at zero coupling:
\begin{eqnarray}
\notag
\texttt{tree} =\sum_{J,J'} \,\langle \texttt{bot}_{J'} |   \big(\mathbb{I} \otimes |v\rangle \langle v |  \big)  | \texttt{top}_{J} \rangle= \frac{w}{(w-1)}\,.
\end{eqnarray} 
Similarly, for the six-point checks we derived the null \texttt{hexagon} starting from the planar frame
\begin{eqnarray}
\notag
&&x_1=(z,\bar z)\,,\, x_2 =(0,0)\,,\, x_3 =(u,\bar u)\,,\\ &&x_4 =(1,1)\,,\,x_5= (w,\bar w)\,,\, x_6=(\infty,\infty)\,, \notag
\end{eqnarray}
and sending three edges on the light-cone $\bar z \to 0\,,\,\bar u \to 1\,,\,\bar w \to \infty$. 
For configurations $a)$ and $b)$ in figure \ref{hexagonBridges},
\begin{eqnarray}
\notag
\texttt{tree}_{a)} = \frac{w}{u(w-1)}\,,\,\,\,\,\,\texttt{tree}_{b)} =\frac{w}{u(w-1)} \left(\frac{w}{w-z}\right)^{h}\,.
\end{eqnarray}

\section{C. Symbols}
As explained in the main text
\begin{eqnarray}\nonumber
\mathcal{F}_{h,k} = \sum_{L=0}^\infty \sum_{n} (g^2 \log\bar w)^n (g^2 \log1/\bar z)^{L-n} \times 
\\
\times \sum_{i} c_{i_1,\dots,i_L}^{(n)}  a_{i_1} \otimes \dots \otimes a_{i_L} \label{symbolAnsatz}
\end{eqnarray}
where the letters $a_i$ take values in the set (\ref{symbolSet}). There are 9 letters in the alphabet (\ref{symbolAnsatz}) so at each loop order $L$, for a given monomial of the stampede logs (i.e. for each $n$) we have in principle $9^L$ words. That is a big number ($5$ is the maximum we have gone to; $9^5=59059$). 
Fortunately there are several simplifications -- some rigorous and some empirically observed -- which allow us to dramatically reduce this number and carry out this procedure up to very high loop order. First not every collections of words in a symbol corresponds to a function. This is guaranteed by the so-called integrability condition which is applied to any two consecutive slots of the symbol, see e.g. \cite{Volovich,Gaiotto:2011dt}. Moreover the logarithmic divergence of a symbol - captured by the first letter - must differ from blue logs, which are factorized out and multiply the symbol as in \eqref{symbol11}. Hence the first letter is chosen from a limited subset which for example in the channel $x_{12}^2=x_{45}^2=0$ is $\{1-1/w, 1-z/w , 1-z/v\}$.

\section{D. Five Loops $\texttt{pentagon}_{1,1}$}
\vspace*{-5mm}
\scriptsize
\begin{align*}
&\mathbb{P}_{1,1}= 1+ \left(t_1^2+t_2^2+t_3^2+t_4^2+t_5^2\right)+\frac{1}{4} \left(t_1^4 \!+ 4 t_1^2 t_2^2 + 2 t_2^4 + 4 t_1^2 t_3^2 + \right.\\
&+4 t_2^2 t_3^2 +\! t_3^4 +  4 t_1^2 t_4^2 \!+4 t_2^2 t_4^2+ 2 t_3^2 t_4^2 + t_4^4 + 2 t_1^2 t_5^2 + 4 t_2^2 t_5^2 + 4 t_3^2 t_5^2 \\
& \left. + 4 t_4^2 t_5^2 + t_5^4\right)+\frac{1}{36}\left(t_1^6 + 9 t_1^4 t_2^2 + \!15 t_1^2 t_2^4 + 5 t_2^6 + 9 t_1^4 t_3^2+27 t_1^2 t_2^2 t_3^2 + \right. \\
& + 15 t_2^4 t_3^2 + 9 t_1^2 t_3^4+ 18 t_2^4 t_4^2 +  9 t_2^2 t_5^4 + 9 t_3^2 t_5^4 + 9 t_4^2 t_5^4 + t_5^6+9 t_3^4 t_5^2 + \\
& + \!9 t_2^2 t_3^4 + t_3^6 + \!9 t_1^4 t_4^2 + 36 t_1^2 t_2^2 t_4^2 + \!18 t_1^2 t_3^2 t_4^2 + 18 t_2^2 t_3^2 t_4^2+ 3 t_3^4 t_4^2 + 9 t_1^2 t_4^4+\\
&   + 9 t_2^2 t_4^4 + 3 t_3^2 t_4^4+t_4^6 + 3 t_1^4 t_5^2 + 18 t_1^2 t_2^2 t_5^2 + 18 t_2^4 t_5^2 +18 t_1^2 t_3^2 t_5^2+ 3 t_1^2 t_5^4 +  \\
&\left.  +18 t_1^2 t_4^2 t_5^2 +  36 t_2^2 t_4^2 t_5^2 + 18 t_3^2 t_4^2 t_5^2 + 9 t_4^4 t_5^2  +  36 t_2^2 t_3^2 t_5^2 \right)+\\
&+ \frac{1}{576}\left(t_1^8 + 16 t_1^6 t_2^2 + 54 t_1^4 t_2^4 + 56 t_1^2 t_2^6 + 14 t_2^8 + 16 t_1^6 t_3^2 + 96 t_1^4 t_2^2 t_3^2 +  \right.\\
&+ 144 t_1^2 t_2^4 t_3^2 +  56 t_2^6 t_3^2 + 36 t_1^4 t_3^4 +96 t_1^2 t_2^2 t_3^4 +  54 t_2^4 t_3^4 +  16 t_1^2 t_3^6 + 16 t_2^2 t_3^6 +\\
&+ 16 t_1^6 t_4^2 + 144 t_1^4 t_2^2 t_4^2  + 240 t_1^2 t_2^4 t_4^2+ 80 t_2^6 t_4^2 +  72 t_1^4 t_3^2 t_4^2 + 216 t_1^2 t_2^2 t_3^2 t_4^2 + \\
 &+ t_3^8 +48 t_1^2 t_3^4 t_4^2 + 48 t_2^2 t_3^4 t_4^2+ 4 t_3^6 t_4^2 + 36 t_1^4 t_4^4 +  144 t_1^2 t_2^2 t_4^4+72 t_2^4 t_4^4 + \\
 & + 48 t_1^2 t_3^2 t_4^4 + 48 t_2^2 t_3^2 t_4^4 + 6 t_3^4 t_4^4 + 16 t_1^2 t_4^6+  4 t_3^2 t_4^6+ 4 t_1^6 t_5^2 + 48 t_1^4 t_2^2 t_5^2   +\\
 & + t_4^8 +  120 t_1^2 t_2^4 t5^2+80 t_2^6 t_5^2  + 48 t_1^4 t_3^2 t_5^2 +  216 t_1^2 t_2^2 t_3^2 t_5^2 + 240 t_2^4 t_3^2 t_5^2 + \\
 &+72 t_1^2 t_3^4 t_5^2 + 144 t_2^2 t_3^4 t_5^2 + 16 t_3^6 t_5^2 + 48 t_1^4 t_4^2 t_5^2 +  288 t_1^2 t_2^2 t_4^2 t_5^2 + 16 t_2^2 t_4^6+ \\
 &+144 t_1^2 t_3^2 t_4^2 t_5^2 + 288 t_2^2 t_3^2 t_4^2 t_5^2 +  48 t_3^4 t_4^2 t_5^2 + 72 t_1^2 t_4^4 t_5^2 + 144 t_2^2 t_4^4 t_5^2+ \\
 & + 48 t_3^2 t_4^4 t_5^2 + 16 t_4^6 t_5^2 + 6 t_1^4 t_5^4 + 48 t_1^2 t_2^2 t_5^4+  72 t_2^4 t_5^4 +48 t_1^2 t_3^2 t_5^4 + \\
 & + 144 t_2^2 t_3^2 t_5^4 +  36 t_3^4 t_5^4 + 48 t_1^2 t_4^2 t_5^4+144 t_2^2 t_4^2 t_5^4 + 72 t_3^2 t_4^2 t_5^4 + 4 t_1^2 t_5^6 +  \\
 &\left. + 36 t_4^4 t_5^4 + 16 t_2^2 t_5^6+ 120 t_2^4 t_3^2 t_4^2 + 288 t_2^4 t_4^2 t_5^2 + 16 t_3^2 t_5^6 + 16 t_4^2 t_5^6 + t_5^8 \right)+\\
 &+ \frac{1}{14400}\left( 25 t_1^8 t_2^2 + 140 t_1^6 t_2^4 + 280 t_1^4 t_2^6 + 210 t_1^2 t_2^8+ 42 t_2^{10}+ 25 t_1^8 t_3^2 + \right.\\
 & + t_1^{10} +250 t_1^6 t_2^2 t_3^2 + 700 t_1^4 t_2^4 t_3^2 + 700 t_1^2 t_2^6 t_3^2+ 210 t_2^8 t_3^2 + 100 t_1^6 t_3^4 +\\
 & + 500 t_1^4 t_2^2 t_3^4 + 700 t_1^2 t_2^4 t_3^4 +  280 t_2^6 t_3^4+ 100 t_1^4 t_3^6 + 250 t_1^2 t_2^2 t_3^6 +\\
 & + 140 t_2^4 t_3^6 +  25 t_1^2 t_3^8 + 25 t_2^2 t_3^8+ 25 t_1^8 t_4^2 + 400 t_1^6 t_2^2 t_4^2 + 1350 t_1^4 t_2^4 t_4^2+\\
 &  + 1400 t_1^2 t_2^6 t_4^2+ t_3^{10}  + 200 t_1^6 t_3^2 t_4^2 + 1200 t_1^4 t_2^2 t_3^2 t_4^2 + 1800 t_1^2 t_2^4 t_3^2 t_4^2 \\
& + 700 t_2^6 t_3^2 t_4^2  + 300 t_1^4 t_3^4 t_4^2+ 800 t_1^2 t_2^2 t_3^4 t_4^2 + 450 t_2^4 t_3^4 t_4^2 + 100 t_1^2 t_3^6 t_4^2 + \\
 &+ 100 t_2^2 t_3^6 t_4^2 +5 t_3^8 t_4^2 + 100 t_1^6 t_4^4 +  900 t_1^4 t_2^2 t_4^4 + 1500 t_1^2 t_2^4 t_4^4 + 500 t_2^6 t_4^4 + \\
&+  300 t_1^4 t_3^2 t_4^4 + 900 t_1^2 t_2^2 t_3^2 t_4^4 +500 t_2^4 t_3^2 t4^4  + 150 t_1^2 t_3^4 t_4^4 + 150 t_2^2 t_3^4 t_4^4 + \\
&+ 10 t_3^6 t_4^4+ 100 t_1^4 t_4^6 + 400 t_1^2 t_2^2 t_4^6 + 200 t_2^4 t_4^6 + 100 t_1^2 t_3^2 t_4^6 + 100 t_2^2 t_3^2 t_4^6 + \\
&+10 t_3^4 t_4^6+ 25 t_1^2 t_4^8  + 25 t_2^2 t_4^8 + 5 t_3^2 t_4^8 + t_4^{10} + 5 t_1^8 t_5^2 +  100 t_1^6 t_2^2 t_5^2 + \\
&+450 t_1^4 t_2^4 t_5^2+ 700 t_1^2 t_2^6 t_5^2 + 350 t_2^8 t_5^2 + 100 t_1^6 t_3^2 t_5^2 + 800 t_1^4 t_2^2 t_3^2 t_5^2 +\\
 &+ 1800 t_1^2 t_2^4 t_3^2 t_5^2 + 1400 t_2^6 t_3^2 t_5^2 + 300 t_1^4 t_3^4 t_5^2 + 1200 t_1^2 t_2^2 t_3^4 t_5^2 +25 t_3^8 t_5^2 + \\
 &+ 200 t_1^2 t_3^6 t_5^2 + 400 t_2^2 t_3^6 t_5^2+1350 t_2^4 t_3^4 t_5^2+ 100 t_1^6 t_4^2 t_5^2 + 1200 t_1^4 t_2^2 t_4^2 t_5^2 + \\
 &+3000 t_1^2 t_2^4 t_4^2 t_5^2 + 2000 t_2^6 t_4^2 t_5^2  +600 t_1^4 t_3^2 t_4^2 t_5^2+ 3000 t_2^4 t_3^2 t_4^2 t_5^2 + \\
 &+1200 t_1^2 t_2^2 t_4^2 t_5^4+ 600 t_1^2 t_3^4 t_4^2 t_5^2+1200 t_2^2 t_3^4 t_4^2 t_5^2 + 100 t_3^6 t_4^2 t_5^2 + \\
 &+ 300 t_1^4 t_4^4 t_5^2 +1800 t_1^2 t_2^2 t_4^4 t_5^2 + 1800 t_2^4 t_4^4 t_5^2 + 600 t_1^2 t_3^2 t_4^4 t_5^2 + 350 t_2^8 t_4^2 +\\
 &+ 1200 t_2^2 t_3^2 t_4^4 t_5^2 + 150 t_3^4 t_4^4 t_5^2 + 200 t_1^2 t_4^6 t_5^2+ 400 t_2^2 t_4^6 t_5^2 +100 t_3^2 t_4^6 t_5^2  +\\
 &+25 t_4^8 t_5^2+\! 10 t_1^6 t_5^4 + 50 t_1^4 t_2^2 t_5^4 +500 t_1^2 t_2^4 t_5^4 + 500 t_2^6 t_5^4 + 150 t_1^4 t_3^2 t_5^4 + \\
 &+900 t_1^2 t_2^2 t_3^2 t_5^4 + 1500 t_2^4 t_3^2 t_5^4+ 300 t_1^2 t_3^4 t_5^4 + 900 t_2^2 t_3^4 t_5^4 + 100 t_3^6 t_5^4 + \\
&+ 150 t_1^4 t_4^2 t_5^4 +1800 t_2^4 t_4^2 t_5^4 + 600 t_1^2 t_3^2 t_4^2 t_5^4+ 1800 t_2^2 t_3^2 t_4^2 t_5^4 + 300 t_1^2 t_4^4 t_5^4 +\\
& +  900 t_2^2 t_4^4 t_5^4 + 300 t_3^2 t_4^4 t_5^4 + 100 t_4^6 t_5^4+ 100 t_1^2 t_2^2 t_5^6 + 200 t_2^4 t_5^6 + 100 t_1^2 t_3^2 t_5^6 + \\
&+400 t_2^2 t_3^2 t_5^6 + 100 t_3^4 t_5^6 + 100 t_1^2 t_4^2 t_5^6 + 400 t_2^2 t_4^2 t_5^6 + 200 t3^2 t_4^2 t_5^6+5 t_1^2 t_5^8+\\
&+ 300 t_3^4 t_4^2 t_5^4 100 t_4^4 t_5^6 + 10 t_1^4 t_5^6 + 25 t_2^2 t_5^8 + 25 t_3^2 t_5^8 +  25 t_4^2 t_5^8 + t_5^{10}+ \\&\left. {\color{magenta} +2700 t_1^2 t_2^2 t_3^2 t_4^2 t_5^2} 
  \right)
  \,.
\end{align*}
\normalsize
We can fix higher order via the bootstrap procedure below~(\ref{startBootstrap}). We could also compute it with the stampedes. With these we access the result with one of the times set to zero so we would miss the term in magenta and more at higher loops. The bootstrap captures them all.
\section{E. Reductions \& Matrix Integrals}
The functions $\mathbb{P}_{h,k}$ enjoy simple reductions when only $t_i,t_{i+1}$-s are non-zero, that is when $x_{i,i+1}^2$ approaches the light-cone faster than the other edges. Besides those following from \eqref{pentared}, other non-trivial reductions read
\begin{align}
\begin{aligned}
\notag
&\mathbb{P}_{h,k}(t_1,t_2,0,0,0) = \tau_{h,k} \Big(\sqrt{t_1^2+t_2^2}\,,\, t_2\Big)\,,
\end{aligned}
\end{align}
where the function $\tau_{h,k}$ is defined in terms of modified Bessel functions of the first kind $I_j(z)$ as
\begin{equation}
\label{tauTwist}
 \tau_{h,k}(x,y)=\det_{1\leq i,j\leq {h+k}} \begin{vmatrix} z_i^{i-j} I_{i-j}\left(2 z_i \right)\end{vmatrix}\,,
\end{equation} 
where $z_i = x$ if $ i\leq h$ and $z_i = y$ if $i > h$. The functions~\eqref{tauTwist} are solutions to a hierarchy of Toda equations 
\begin{equation}
{\tau_{h,k}^2} \mathcal D_{x} \left(\mathcal D_{x}+\mathcal D_{y}\right) \log \tau_{h,k}= 4 x^2 \, {\tau_{h+1,k} \tau_{h-1,k}}\,,
\end{equation}
that is the reduction of the first of \eqref{Todas} when the cusp time $t_3$ is switched off. 
The seed $ \tau_{0,k}(x,y) = \tau_{k}(y)$ counts random walks inside the Weyl chamber of $\mathbb{Z}^{\ell}$~\cite{GZ}. 
Formula \eqref{tauTwist} can be further generalized introducing different times for each row of the matrix in \cite{general_det}
\begin{equation}
\label{tauTwist_gen}
 \tau(z_1,\dots,z_{\ell})=\det_{1\leq i,j\leq {\ell}} \begin{vmatrix} z_i^{i-j} I_{i-j}\left(2 z_i \right)\end{vmatrix}\, .
\end{equation}
Using the integral representation for the Bessel functions $
I_m \sim \int e^{i \theta m +z \cos \theta }\,,$
 we can re-write \eqref{tauTwist_gen} as an integral over $\ell$ angles $\theta_k$
 \begin{equation}
 \notag
 \tau(z_1,\dots,z_{\ell})\!=  \!\int\limits_0^{2\pi}\! \!\Big(\prod_k \frac{d\theta_k}{2\pi} e^{2 z_k  \cos \theta_k}\Big) \det_{m,n} \begin{vmatrix} (z_m e^{i\theta_m})^{m-n} \end{vmatrix}\,,
\end{equation}
The integrated determinant can be simplified collecting the Vandermonde of a matrix $Z$ with eigenvalues $\{z_k e^{i\theta_k}\}$
\begin{eqnarray*}
 \det_{m,n} \begin{vmatrix} (z_m e^{i\theta_m})^{m-n} \end{vmatrix} \,\,\,&&=   \Delta(Z)\prod_{k}(z_k e^{i\theta_k})^{1-k}\,.
\end{eqnarray*}
In fact, introducing a set of auxiliary times $\{y_k\}$ and the following differential operator
\begin{equation}
\mathfrak{D}_{y,z} = \prod_{k=1}^{\ell} \frac{1}{(k-1)!} \Big(\frac{\partial_{y_k}}{z_k}\Big)^{k-1} \Big|_{y_k=0} \,,
\end{equation}
we can relate the function \eqref{tauTwist_gen} to a nice matrix integral
\begin{equation*}
\tau_{\ell}(z_1,\dots,z_{\ell})
 = \mathfrak{D}_{y,z} \int \!\!\Big(\prod_k \frac{d\theta_k}{2\pi}\Big) \Delta(Z)\Delta(Y^{\dagger}) e^{\text{Tr}(Z+Z^{\dagger})}.
\end{equation*}
The integral in the rhs is reminiscent of normal matrix models with angular integration (see the nice review \cite{vanMoerbeke} and references therein); when $y_k=z_k$ it coincides with the Brezin-Gross unitary matrix model \cite{Brezin}. 
For the special reduction $z_i=z_j$, \eqref{tauTwist_gen} matches with the notorious plaquette matrix model of \cite{GWW}.

\bibliographystyle{apsrev4-1}

\begin{thebibliography}{0}
\bibitem{generalLC}
These three simplifications hold as we take the full null limit of the pentagon. They can easily be relaxed following \cite{stampedes1} to describe a more general null limit where only $\bar w$ and $1/\bar z$ are sent to zero as discussed in the conclusions and in appendix B.
\bibitem{hexa_hop}
Note that for even sided polygons there is no sub-leading cusp time in the stampede, as each cusp is extremal of an edge which is expanded in light-like direction; in this sense, the hexagon checks are even cleaner than the pentagon ones.
\bibitem{general_det}
While this is certainly an interesting generalization from a matrix integral point of view, we do not know if it corresponds to a particular null polygon with a particular bridge setup; probably not.
\end{thebibliography}

\end{document}